\documentclass{iau307}
\usepackage{graphicx}
\usepackage{natbib}
\usepackage{url}
\usepackage{dtklogos}
\bibpunct{(}{)}{;}{a}{}{,}

\title[The BOB Survey] 
{The B Fields in OB Stars (BOB) Survey}

\author[T. Morel, et al.]   
{T. Morel,$^{1}$
N. Castro,$^{2}$
L. Fossati,$^{2}$
S. Hubrig,$^{3}$
N. Langer,$^{2}$
N. Przybilla,$^{4}$
M. Sch\"oller,$^{5}$
T. Carroll,$^{3}$
I. Ilyin,$^{3}$
A. Irrgang,$^{6}$
L. Oskinova,$^{7}$
F. R. N. Schneider,$^{2}$
S. Simon D\'{\i}az,$^{8,9}$
M. Briquet,$^{1}$
J. F. Gonz\'alez,$^{10}$
N. Kharchenko,$^{11}$
M.-F. Nieva,$^{4,6}$
R.-D. Scholz,$^{3}$
A. de Koter,$^{12,13}$
W.-R. Hamann,$^{7}$
A. Herrero,$^{8,9}$
J. Ma\'{\i}z Apell\'aniz,$^{14}$
H. Sana,$^{15}$
R. Arlt,$^{3}$
R. Barb\'a,$^{16}$
P. Dufton,$^{17}$
A. Kholtygin,$^{18}$
G. Mathys,$^{19}$
A. Piskunov,$^{20}$
A. Reisenegger,$^{21}$
H. Spruit,$^{22}$
\and S.-C. Yoon$^{23}$
}

\affiliation{$^{1}$Institut d'Astrophysique et de G\'eophysique, Li\`ege, Belgium \\ email: {\tt morel@astro.ulg.ac.be} \\[\affilskip]
$^{2}$Argelander-Institut f\"ur Astronomie, Bonn, Germany  \\[\affilskip]
$^{3}$Leibniz-Institut f\"ur Astrophysik Potsdam (AIP), Potsdam, Germany  \\[\affilskip]
$^{4}$Institute for Astro- and Particle Physics, University of Innsbruck, Austria  \\[\affilskip]
$^{5}$European Southern Observatory, Garching, Germany  \\[\affilskip]
$^{6}$Dr. Remeis Observatory \& ECAP, Bamberg, Germany  \\[\affilskip]
$^{7}$Institut f\"ur Physik und Astronomie der Universit\"at Potsdam, Germany  \\[\affilskip]
$^{8}$Instituto de Astrof\'{\i}sica de Canarias, La Laguna, Spain  \\[\affilskip]
$^{9}$Universidad de La Laguna, Dpto. de Astrof\'{\i}sica, La Laguna, Spain  \\[\affilskip]
$^{10}$Instituto de Ciencias Astronomicas, de la Tierra, y del Espacio (ICATE), San Juan, Argentina  \\[\affilskip]
$^{11}$Main Astronomical Observatory, Kiev, Ukraine \\[\affilskip]
$^{12}$Astronomical Institute Anton Pannekoek, Amsterdam, The Netherlands \\[\affilskip]
$^{13}$Instituut voor Sterrenkunde, Universiteit Leuven, Leuven, Belgium\\[\affilskip]
$^{14}$Instituto de Astrof\'{\i}sica de Andaluc\'{\i}a-CSIC, Granada, Spain \\[\affilskip]
$^{15}$ESA/Space Telescope Science Institute, Baltimore, USA \\[\affilskip]
$^{16}$Departamento de F\'{\i}sica, La Serena, Chile \\[\affilskip]
$^{17}$Astrophysics Research Centre, Belfast, UK \\[\affilskip]
$^{18}$Chair of Astronomy, Saint-Petersburg State University, Russia \\[\affilskip]
$^{19}$European Southern Observatory, Santiago, Chile \\[\affilskip]
$^{20}$Institute of Astronomy of the Russian Acad. Sci., Moscow, Russia \\[\affilskip]
$^{21}$Pontificia Universidad Cat\'olica de Chile, Santiago, Chile \\[\affilskip]
$^{22}$Max-Planck-Institut f\"ur Astrophysik, Garching, Germany \\[\affilskip]
$^{23}$Department of Physics and Astronomy, Seoul National University, Seoul, Republic of Korea
}

\pubyear{2014}
\volume{307} 
\pagerange{}
\jname{New windows on massive stars: asteroseismology, interferometry, and spectropolarimetry}
\editors{G. Meynet, C. Georgy, J.H. Groh \& Ph. Stee, eds.}

\begin{document}

\maketitle

\begin{abstract}
The B fields in OB stars (BOB) survey is an ESO large programme collecting spectropolarimetric observations for a large number of early-type stars in order to study the occurrence rate, properties, and ultimately the origin of magnetic fields in massive stars. As of July 2014, a total of 98 objects were observed over 20 nights with FORS2 and HARPSpol. Our preliminary results indicate that the fraction of magnetic OB stars with an organised, detectable field is low. This conclusion, now independently reached by two different surveys, has profound implications for any theoretical model attempting to explain the field formation in these objects. We discuss in this contribution some important issues addressed by our observations (e.g., the lower bound of the field strength) and the discovery of some remarkable objects.
\end{abstract}

\section{Magnetic fields in OB stars}
Magnetic fields are one of the key factors affecting the evolution and properties of massive stars. Yet it is only very recently that the number of magnetic OB stars known has reached a level that allows us to evaluate the field incidence, examine the properties of the fields, and critically test the various models that were proposed for their creation. The picture now emerging is that relatively strong fields (above, say, 100-200 G at the surface) are only found in a small fraction of all massive stars and that the field topology is rather simple (dipolar, or, in some rare cases, low-order multipolar). Moreover, the field strength is not directly linked to the stellar parameters. These characteristics are similar to those presented by the chemically-peculiar Ap/Bp stars \citep{donati09}. This suggests a similar origin of the field. 

Despite the dramatic progress made over the last few years, the answers to some important questions are still eluding us, e.g., the effects of fields on the internal rotational profile and on the transport of the chemical species. Even how the field is created is not completely settled. The magnetic field permeating the interstellar medium (ISM) is amplified during star formation and may naturally relax into a large-scale, mostly poloidal field emerging at the surface \citep[e.g.,][]{braithwaite04}. The similarity between the magnetic properties of OB and Ap/Bp stars suggests that today we observe the remnant of the field frozen-in in the ISM. However, in view of the significant fraction of OB stars that may suffer a merger or a mass-transfer event during their evolution \citep{sana12}, it cannot be ruled out that fields are created through such processes \citep[e.g.,][]{wickramasinghe14}. Finally, a dynamo operating in subsurface convection layers could produce short-lived, spatially localised magnetic structures \citep{cantiello11} that are, however, much more challenging to detect. 

A better understanding of the origin and effects of magnetic fields requires the identification of additional magnetic objects (e.g., only in about ten O stars has a field been firmly detected). This has motivated us to initiate the B fields in OB stars (BOB) survey.

\section{The BOB survey}
A total of 35.5 observing nights were allocated during P91-P96 as part of an ESO large programme (191.D-0255; PI: Morel). About 20 nights are dedicated to obtain snapshot observations of a large number of OB stars, while the remaining nights are devoted to confirm the field detection for the candidate magnetic stars and to better characterise the field properties for those that are firmly identified as being magnetic. Two different state-of-the-art instruments with circular polarisation capabilities were used (with low and high spectral resolution, respectively): FORS2 at the VLT for the fainter targets and HARPSpol at the 3.6-m telescope at La Silla for the brighter ones. About two thirds of the nights are scheduled on HARPSpol. As of July 2014, 20 nights of observations were completed. Only one night (with FORS2) was lost because of bad weather. 

Previously known magnetic OB stars on average appear to have rotation speeds significantly lower than the rest of the population. We therefore mostly targeted stars with $v\sin i$ below 60 km s$^{-1}$ to enhance the probability of detecting fields. Contrary to the MiMeS survey (Grunhut et al., these proceedings), we concentrate on normal, main-sequence OB stars and do not consider, e.g., Of?p, Be or Wolf-Rayet stars. The sample is composed in roughly equal parts of O ($\sim$40\%) and B ($\sim$60\%) stars. The vast majority are late O- and early B-type stars. BOB and MiMeS can be viewed as two complementary surveys in the sense that there are very few targets in common.

One important aspect of our survey is that the data reduction and analysis are carried out completely independently by two groups (one from the Argelander-Institut f\"ur Astronomie in Bonn and the other from the Leibniz-Institut f\"ur Astrophysik in Potsdam) to ensure that the results are robust. The two groups process both the FORS2 and HARPSpol data separately, and employ different tools and analysis techniques \citep[for details, see][]{hubrig14}.

\section{The occurrence of magnetic fields in massive stars}
The results obtained by the only other large-scale survey of this kind up to now \citep[MiMeS;][]{wade13} indicate that about 7\% of massive stars host a magnetic field detectable with current instrumentation ($\gtrsim$ 100 G). We have so far observed 98 targets and only unambiguously detected five magnetic stars. For all the stars, the detection is not only confirmed by the two groups (Bonn and Potsdam), but the field measurements also systematically agree within the errors. In addition, the field is detected at a high significance level with both FORS2 and HARPSpol.

Therefore, our results tend to support those independently obtained by MiMeS, and confirm that the incidence rate of strong magnetic fields is low in massive stars and is similar to that inferred for intermediate-mass stars. It should be emphasised, however, that a number of candidate magnetic stars are still being followed up and that the preliminary incidence rate that we obtain ($\sim$5\%) may eventually be revised upwards. 

Regardless of the exact figures, the scarcity of strong fields has far-reaching implications, from the interpretation of the statistical properties of young stellar populations (e.g., impact of magnetic braking on the rotational velocities, X-ray characteristics) to the fate of massive stars as degenerate objects following the supernova explosion (e.g., as magnetars). 

\section{The first magnetic stars discovered in the course of the survey}
\subsection{A magnetic field in a multiple system in the Trifid nebula}
One of the aims of our survey is to uncover magnetic stars with specific and unusual characteristics that would allow us to discriminate between the various channels that could lead to the field formation. An interesting discovery is therefore the detection of a magnetic field in a multiple system in the Trifid Nebula \citep{hubrig14}, which is a very young and active site of star formation. We first observed the three brightest components in the central part of the nebula \citep[A, C, and D;][]{kohoutek99} with FORS2 and clearly detected a circularly polarised signal in component C (HD 164492C). 

Further observations on two consecutive nights with HARPSpol confirmed the existence of a field with a longitudinal strength ranging from 400 to 700 G. These high-resolution observations reveal complex and variable line profiles pointing towards a multiple system (made up of at least two early B-type stars). The situation is complicated further by the possible existence of chemical patches on the surface of some components. We will keep monitoring this system to establish whether only one or more components are magnetic. A complete characterisation of this peculiar system may provide valuable information about the interplay between binarity and magnetic fields in massive stars.

\subsection{A new magnetic, helium-rich star with a tight age constraint}
The variability of the rare magnetic, helium-rich stars (the prototype is $\sigma$ Ori E) arises from a dipolar field that is tilted with respect to the rotational axis. The competition between radiative levitation and gravitational settling in the presence of a stellar wind leads to photospheric abundance anomalies. Some stars undergo rapid rotational braking \cite[e.g.,][]{mikulasek08}, which provides an unique opportunity to study virtually in real time the poorly-understood effects of angular momentum loss through magnetically channeled, line-driven winds \cite[e.g.,][]{ud_doula09}. 

Our observations of the B1 star CPD --57$^\circ$ 3509 in the young ($\sim$10 Myr) open cluster NGC 3293 with FORS2 and HARPSpol reveal a strong and rapidly varying field (by up to 900 G for the longitudinal component between two consecutive nights). The field is found to change polarity, which shows that both magnetic hemispheres are visible as the star rotates. The polar field exceeds 3 kG assuming a dipole geometry. A preliminary NLTE spectral analysis indicates that the star is helium rich ($\sim$3 times solar) and has evolved throughout about one third of its main-sequence lifetime (Przybilla et al., in prep.). This is one of the most evolved He-rich stars with a tight age constraint, promising to provide crucial information on the evolution of stars with magnetically-confined stellar winds. 

\subsection{A non-peculiar magnetic O star}
\begin{figure}[t!]
\begin{center}
\includegraphics[trim=0cm 8.9cm 0cm 7.5cm, clip=true, width=0.65\textwidth]{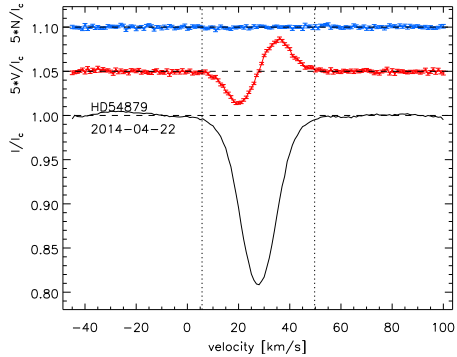} 
\caption{Stokes $I$ ({\it black}), $V$ ({\it red}), and diagnostic null $N$ ({\it blue}) profiles of HD 54879 obtained with HARPSpol through least-squares deconvolution (LSD).  From Castro et al., in preparation.}
\label{fig_hd54879}
\end{center}
\end{figure}

The few magnetic O stars known are very often peculiar. Their strong magnetic fields are believed to give rise to spectral peculiarities and/or to drive periodic line-profile variations (e.g., the Of?p stars or $\theta^1$ Ori C). In contrast, we have discovered a narrow-lined O9.7 V star (HD 54879) hosting a strong field (with a dipole strength above 2 kG; Fig.\ref{fig_hd54879}), yet displaying no evidence in the few optical spectra taken over five years for spectral peculiarity or variability. Only the broad and emission-like H$\alpha$ profile is variable. This might be related to the presence of a centrifugal magnetosphere (Castro et al., in prep.). Further observations are necessary to confirm the lack of spectral peculiarities and, if so, to understand the distinct behaviour with respect to other strongly magnetised O stars. A parallel investigation of the magnetic variability also needs to be undertaken.

\subsection{Weak fields in OB stars}
One of the most intriguing properties of magnetic stars of intermediate mass is the bimodal nature of the fields that are either strong (above 300 G) or extremely weak ($\lesssim$1 G). The lack of objects with ordered fields of intermediate strength appears to reveal a real dichotomy \citep[e.g.,][]{lignieres14}. Investigating the origin of this ``magnetic desert'' is essential to understand the origin and evolution of fields in stars that cannot support a dynamo acting in a deep, outer convective envelope.  

\begin{figure}[t!]
\begin{center}
\includegraphics[trim=0cm 9.5cm 0cm 7.4cm, clip=true, width=0.9\textwidth]{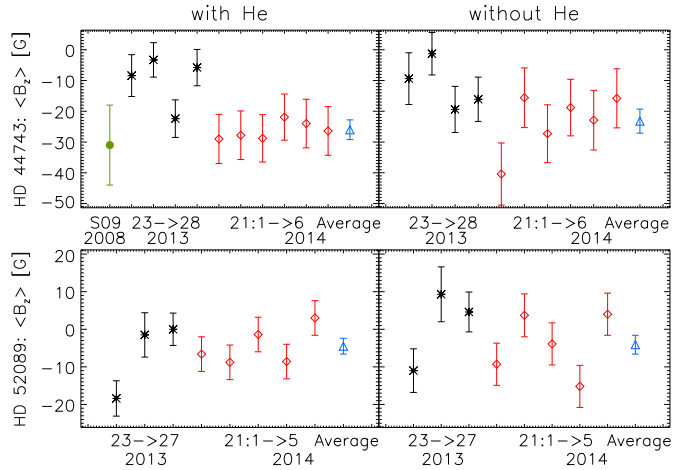} 
\caption{Time series of the longitudinal field measurements of $\beta$ CMa (top panels) and $\varepsilon$ CMa (bottom panels) using a line mask which contains (left panels) or does not contain (right panels) helium lines. Black crosses: observations carried out on four different nights in December 2013. Red rhombs: values obtained from consecutive observations on 21 April 2014 (the blue triangle is the average value obtained on that night). Green circle: ESPaDOnS measurement of $\beta$ CMa carried out in 2008 by \citet{silvester09}. From Fossati et al., in preparation.}
\label{fig_weak_fields}
\end{center}
\end{figure}

To estimate the lower bound of the field in more massive stars, we have obtained very high-quality observations with HARPSpol of two very bright, early B-type stars ($\beta$ CMa and $\varepsilon$ CMa), and repeatedly detected for both stars a weak Zeeman signature across the line profiles. Interestingly, the field appears to be constant within the errors and relatively weak in both cases. The longitudinal components are at most $\sim$30 G in modulus (Fig.\ref{fig_weak_fields}), which translates into a polar strength of $\sim$150 G assuming a dipolar geometry. Although all the available measurements of $\beta$ CMa are consistent with a field of that magnitude, there is some indication in the literature for a stronger field in $\varepsilon$ CMa \citep{hubrig09, bagnulo12}. Therefore, the case for a weak field is much stronger in $\beta$ CMa.

Detecting fields of that magnitude in fainter targets is very challenging. Are the weak fields found in these two bright objects only the tip of the iceberg? Is there a large population of stars with weaker fields or that are even truly non-magnetic \citep[see][]{neiner14}? It seems conceivable that weak fields are considerably more widespread than the data currently available would suggest. Fields that are remnants of the star formation are prone to decay on evolutionary time-scales \citep[see][]{landstreet08}. A different field strength distribution for intermediate- and high-mass stars thus raises the issue of a mass-dependent time-scale for the field decay (Fossati et al., in prep.).

\begin{acknowledgments}
TM acknowledges financial support from Belspo for contract PRODEX GAIA-DPAC. LF acknowledges financial support from the Alexander von Humboldt Foundation. MB is F.R.S.-FNRS Postdoctoral Researcher, Belgium.
\end{acknowledgments}

\bibliographystyle{iau307}
\bibliography{morel_talk}

\begin{thebibliography}{}

\bibitem[\protect\astroncite{{Bagnulo} et~al.}{2012}]{bagnulo12}
{Bagnulo}, S., {Landstreet}, J.~D., {Fossati}, L., \& {Kochukhov}, O. 2012,
\newblock {\em \aap} 538, A129

\bibitem[\protect\astroncite{{Braithwaite} \& {Spruit}}{2004}]{braithwaite04}
{Braithwaite}, J. \& {Spruit}, H.~C. 2004,
\newblock {\em \nat} 431, 819

\bibitem[\protect\astroncite{{Cantiello} \& {Braithwaite}}{2011}]{cantiello11}
{Cantiello}, M. \& {Braithwaite}, J. 2011,
\newblock {\em \aap} 534, A140

\bibitem[\protect\astroncite{{Donati} \& {Landstreet}}{2009}]{donati09}
{Donati}, J.-F. \& {Landstreet}, J.~D. 2009,
\newblock {\em \araa} 47, 333

\bibitem[\protect\astroncite{{Hubrig} et~al.}{2009}]{hubrig09}
{Hubrig}, S., {Briquet}, M., {De Cat}, P., {et~al.} 2009,
\newblock {\em Astronomische Nachrichten} 330, 317

\bibitem[\protect\astroncite{{Hubrig} et~al.}{2014}]{hubrig14}
{Hubrig}, S., {Fossati}, L., {Carroll}, T.~A., {et~al.} 2014,
\newblock {\em \aap} 564, L10

\bibitem[\protect\astroncite{{Kohoutek} et~al.}{1999}]{kohoutek99}
{Kohoutek}, L., {Mayer}, P., \& {Lorenz}, R. 1999,
\newblock {\em \aaps} 134, 129

\bibitem[\protect\astroncite{{Landstreet} et~al.}{2008}]{landstreet08}
{Landstreet}, J.~D., {Silaj}, J., {Andretta}, V., {et~al.} 2008,
\newblock {\em \aap} 481, 465

\bibitem[\protect\astroncite{{Ligni{\`e}res} et~al.}{2014}]{lignieres14}
{Ligni{\`e}res}, F., {Petit}, P., {Auri{\`e}re}, M., {Wade}, G.~A., \&
  {B{\"o}hm}, T. 2014,
\newblock {\em ArXiv e-prints (1402.5362)}

\bibitem[\protect\astroncite{{Mikul{\'a}{\v s}ek} et~al.}{2008}]{mikulasek08}
{Mikul{\'a}{\v s}ek}, Z., {Krti{\v c}ka}, J., {Henry}, G.~W., {et~al.} 2008,
\newblock {\em \aap} 485, 585

\bibitem[\protect\astroncite{{Neiner} et~al.}{2014}]{neiner14}
{Neiner}, C., {Monin}, D., {Leroy}, B., {Mathis}, S., \& {Bohlender}, D. 2014,
\newblock {\em \aap} 562, A59

\bibitem[\protect\astroncite{{Sana} et~al.}{2012}]{sana12}
{Sana}, H., {de Mink}, S.~E., {de Koter}, A., {et~al.} 2012,
\newblock {\em Science} 337, 444

\bibitem[\protect\astroncite{{Silvester} et~al.}{2009}]{silvester09}
{Silvester}, J., {Neiner}, C., {Henrichs}, H.~F., {et~al.} 2009,
\newblock {\em \mnras} 398, 1505

\bibitem[\protect\astroncite{{ud-Doula} et~al.}{2009}]{ud_doula09}
{ud-Doula}, A., {Owocki}, S.~P., \& {Townsend}, R.~H.~D. 2009,
\newblock {\em \mnras} 392, 1022

\bibitem[\protect\astroncite{{Wade} et~al.}{2013}]{wade13}
{Wade}, G.~A., {Grunhut}, J., {Alecian}, E., {et~al.} 2013,
\newblock {\em ArXiv e-prints (1310.3965)}

\bibitem[\protect\astroncite{{Wickramasinghe} et~al.}{2014}]{wickramasinghe14}
{Wickramasinghe}, D.~T., {Tout}, C.~A., \& {Ferrario}, L. 2014,
\newblock {\em \mnras} 437, 675

\end{thebibliography}

\begin{discussion}

\discuss{Landstreet}{Possible detection of a $\sigma$ Ori E analogue is very interesting. Have you looked for indications of a trapped magnetosphere with detailed comparisons of Balmer line profiles as observed with model profiles, to identify weak emission or shell absorption?}

\discuss{Morel}{This is certainly something we plan to do, but we have still not investigated this in detail. Dedicated follow-up observations are needed to fully characterise the behaviour of the Balmer lines. We have recently been granted time on UVES to monitor this object. These spectra can be used for that purpose.}

\discuss{Wade}{The detection of $V$ signatures in $\beta+\varepsilon$ CMa rely on magnetic precision of $\sim5$ G. For how many other stars is a comparable precision achieved? (Since this will tell us about the frequency of such apparently weak fields.)}

\discuss{Morel}{A comparable precision was only achieved for another star, and in that case a field was not detected. Such high-quality data can only be obtained for very bright stars, and it cannot be ruled out that weak fields also exist in the fainter targets.}

\discuss{Aerts}{Did you try to fold your polarimetric data of $\beta$ CMa on the seismic rotation period to check if they are consistent, as is the case for V2052 Oph and HD 43317?}

\discuss{Morel}{We still have too few measurements to investigate whether the magnetic period is compatible with the rotational one determined by Mazumdar et al. (2006, A\&A, 459, 589). However, it has to be kept in mind that the weakness of the magnetic signal may hamper a clear detection of the variability.}

\end{discussion}

\end{document}